# Determining Positive Cancer Rescue Mutations in p53 Based Cancers by using Artificial Intelligence

Kaan Aygen
The Koc School
gunkaana2018@stu.kocschool.k12.tr

Ismet Berkay Celik
The Koc School
ismetberkayc2018@stu.koc-school.k12.tr

Umut Eser
Harvard Medical School
umuteser@gmail.com

I. INTRODUCTION

A mutation in a protein-coding gene in DNA can alter the protein structure coded by the same gene. Structurally altered proteins usually lose their functions and sometimes gain an undesirable function instead. These types of mutations and their effects can result in genetic diseases or antibiotic resistant bacteria, among other health issues. Important curing methods have been developed by detecting mutations against AIDS as well as genetic diseases ([1]; [2]). Another example is the influenza virus. The reasons why a vaccination developed to fight against influenza does not work the following year are (a) the mutation of its DNA and (b) the outbreak of the virus after it has been mutated especially if it is a virus that escaped the vaccinations target ([3]). Due to such reasons it is highly important to know in advance the location of a potential mutation in a protein as well as the problems it might cause the medical sciences.

Like other diseases mentioned, cancer also appears as a result of a growing cell or a mutation of accumulated proteins that are actually in charge of the controlled death of a cell (apoptosis). The uncontrolled growth of a cell or its uncontrolled division creates a tumor. In a healthy person these minor tumors are constantly created. A wholly functioning p53 protein stops cancer by killing the cells converted into tumors in a controlled way or by preventing the genes, which are in charge of the division, expressing themselves (Figure 1) ([4]; [5]; [6]). A p53 protein, which has such a critical role in preventing cancer, can also be the cause of it when it loses its functions due to a mutation. If the phenotypic functions of these mutations can be foreseen, a groundbreaking cancer cure can be developed with a smart and target oriented medicine design. Although some types of cancer cures are dependent on pharmacological studies, they are usually centered around casting light on the mutational reasons for the cancer.

Predicting the results of a mutation in any amino acid within a protein depends on the estimation of the mutations effect on the protein structure. This is already a very difficult task. It gets even harder combinatorically when multiple mu- tations occur at the same time. For instance, a p53 protein that consists of 393 amino acids (taking into consideration that each amino acid can transform into any one of the 19 amino acids) gives us a wide configuration of approximately $19 \times 10$. In a problem where the possibilities of mutation configurations are increasing exponentially, even if 1000 mutations were tested per day, all the days of the universe would not be sufficient enough. And even if each experiment would cost just 1 cent, all the money in the world would not be enough to test them all. Therefore, we need a sensitive and a strong computational algorithm that has the skills to generalize rules even from a tiny sample.

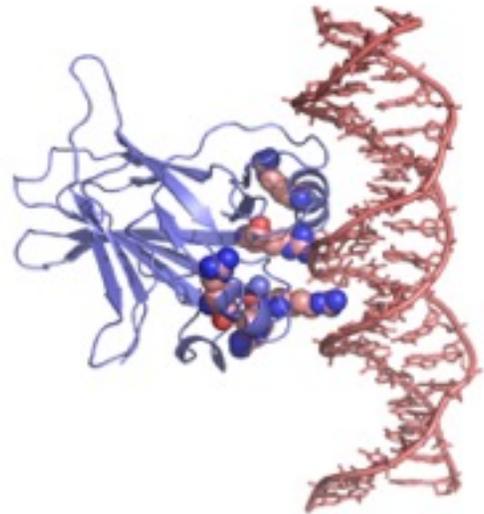

Fig. 1. p53 protein and its amino acids that interact with DNA

In the name of developing computational algorithm, the International Agency for Research on Cancer (IARC) published a TP53 mutation database that contains 31,420 different p53 gene mutations that are diagnosed in cancer patients ([7]; [8]). Approximately 70% of these wide mutation configurations are due to one single amino acid mutation that takes place where p53 binds to DNA. Although it is highly crucial, the complete structure of p53 protein remains undiscovered. We can only create homological models through biophysical simulations since we only know the crystal structure of the functional region of p53 ([9]; [10]). We can also only calculate the effects of a mutation of a protein on a 1 dimensional sequence, a 2 dimensional electronegativity surface profile and in 3 dimensional amino acid locations.

These calculations do not give us enough idea on the functions of a direct mutation since the organization of amino acids in a p53 protein is highly complex and has synergistic relationships. We can briefly define the problem of categorizing the p53 mutations as follows: to foresee whether a p53 protein is active or inactive as a result of alterations in a given series of amino acids.

There have been studies in the field of classic machine learning and statistical techniques ([10]; [11]). However, their performances have so far been far below expectations. One of the important reasons of this low performance is because they have overlooked which amino acids are important in provided data, the approach of heuristic algorithm with evolutional com- parisons, and the interactions in between linear dimensional reduction methods and nonlinear relationships.

Artificial intelligence studies have shown ground breaking developments in recent years thanks to capacity increases in computer technology and their ability to access all sorts of data ([12]). The most important development of all is neural networks models. These models can learn the necessary characteristics from the provided data by themselves in order to foresee the result. This method, known as deep learning, facilitates our lives and provides us with driverless cars, voice recognition systems, visual analysis and unmanned aircrafts among others and resolves our daily problems ([13]). By applying artificial intelligence to medicine and genetics great problems have been overcome in biology in recent years ([14]; [15]; [16]; [17]; [18]).

In this study we have used artificial neural networks, which are one of the latest artificial intelligence technologies, to determine the effects of cancer mutations. The model we developed has given more successful results compared to other methods. We foresee that our model will bring a new dimension to medical research and the medicine industry.

## II. Method

We downloaded data provided by the IARC that consisted of 31,420 different p53 mutations from the California University Irvine archives ([19]). In each data set which has multivariate characteristics, there are 5,409 features for each mutation. 2,510 of these scores feature a 1 dimensional sequence of the mutated protein ([20]), 2,316 score features of 2 dimensional electronegative and the remaining 582 score 3 dimensional position features (Figure 2).

First of all, without altering the order of this data, 1 dimensional, 2 dimensional and 3 dimensional features were recorded separately in a hierarchical data format (hdf5) as developed by NASA for large dimensional data. Since 92% of the data was negative, positive (active) and negative (inactive) sets were not balanced. When we trained our model with an unbalanced set, it only gave negative results. Therefore, during the training of the model we fed it with an equal amount of positive and negative samples. We divided the data into two;
2000 tests and 29,420 training data. The test data model was not used during the training period; it was only used to trial the model.

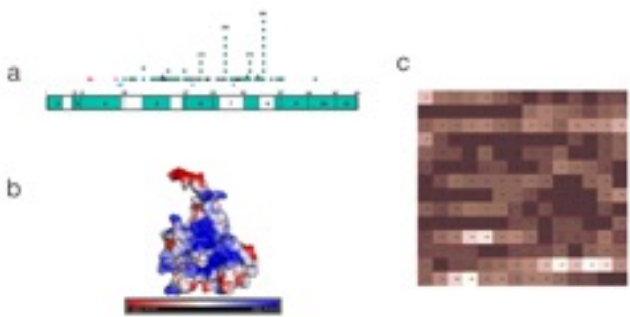

Fig. 2. Characteristics of the input data. (a) 1 dimensional features (Adapted from [20]). (b) 2 dimensional features, blue color indicates positive charge and red color indicates negative charge. (c) 3 dimensional features, boxes indicate 3 dimensional distances to a healthy version of p53 amino acids in a mutational condition. As the box gets whiter it indicates that it is getting further away from its healthy condition (Image adapted from [10]).

We wrote our model in Python language with the Tensor- flow deep learning package ([21]). We connected our model, which has 2 convolutional and 2 complete connection layers, to a 2 neuron output. We determined the activation function of neurons in each layer (except from the final layer) as a rectified linear unit (ReLU), and in the final layer as hyperbolic tangent (tanh). We made it have a possible distribution of a mold total of 1. We did this by filtering the 2-neuron-output through softmax, thus having neuron values in between 0 and 1. We then coded the active mutations as [1,0] and inactive mutations as [0,1]. We determined our loss function as categorical cross entropy, which is widely used for binary categorization, to train the neurons in an artificial neural network model. We evaluated the training of the model by showing the data over and over again and by looking at the changes to loss function over time. Each training time took approximately 10 minutes in a standard laptop.

In order to compare our deep learning model that we adopted from the field of artificial intelligence to other mod- els that have been used until recent years, we utilized the scikitlearn package ([22]). We also applied Support Vector Machines (SVM) and Logistic Regression models.

## III. Findings

First of all, we trained our model individually for a mutated proteins 1 dimensional, 2 dimensional and 3 dimensional features. We later on tried it on test data. The performance alteration over the samples given to the model can be seen in Image 3. If we foresaw it completely randomly, the performance would be 50% as we tested it on an equal amount of positive and negative data. As can be viewed in the image, the final precision levels of predicting mutated protein activity stand at 76%, 89% and 85% respectively, and they are higher compared to random predicting levels. Since our model was developed on the basis of artificial neural networks, we can easily combine different input categories and create a combined model with a higher capacity. When we trained this combined model, we obtained a better performance (98%) than the ones obtained in the single input category (Figure 3). In order to compare our model based on deep learning to other machine learning techniques, we categorized the same data by using Support Vector Machines (SVM) and Logistic Regression from the same scikitlearn package ([22]). We optimized hyper parameters with grid search and reached a level of 73% precision with SVM and 62% precision with Logistic Regression (4a). This proves that our model has a higher success rate compared to others.

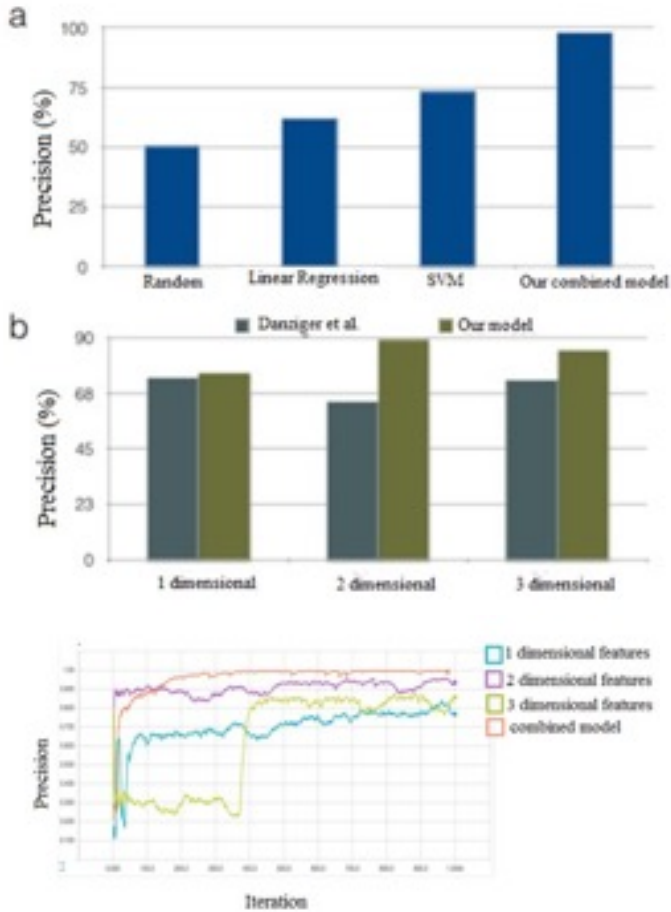

Fig. 3. Performance of trained models of different input data features according to iteration.

Categorizing p53 mutations have been studied but different methods have been used ([10]; [23]). In a similar way, when we compared our model to similar studies we noticed that in each input category our models performance is significantly higher (Figure 4b).

The final layer of our deep learning model has a 2-neuron- output that tells us whether the mutation is active or inactive. The penultimate layer provides a representation of data which is complex and with intricate features. We visualized active and inactive data in order to show that our models input data features, which are in complex interaction among themselves, can be transformed to a comprehendible representation space by Principal Component Analysis (PCA) and t-distributed Stochastic Neighbor Embedding (t-SNE) methods (Figure 5) ([24]). As can be viewed, the models representation layer projected the complex input into a comprehendible space that can easily categorize it.

## IV. RESULTS AND DISCUSSION

p53 protein mutations, which play an instrumental role in various types of cancer and have been studied thousands of times as it is highly important for the medical world to understand them, and what type of results they would produce, could not be foreseen satisfactorily up until now. In this study we applied the artificial neural network, which is one of the latest technologies in artificial intelligence, to p53 mutations in order to foresee problems. We discovered that our model has a higher performance and a higher success rate compared to not only the two other widely used methods, but also to other formerly published academic studies. Since our model is based on artificial neural networks, it can combine different modules in a much easier way, and, in doing, so we can obtain a more developed combined model. We have proved that our combined model is more successful than the single input models.

Fig. 4. Comparison of our model to (a) other common techniques (SVM, Linear Regression) and (b) previous academic studies.

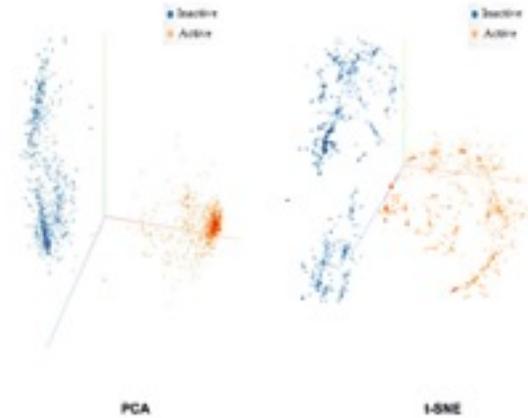

Fig. 5. Visualization of our model by (a) PCA and (b) low dimensional t-SEN in a representational layer. Active and inactive samples that cannot be separated in an input layer can be separated clearly in a representation layer.

Since the number of all p53 mutated types is vast, the active or inactive condition of all mutations cannot be stored in a database. Therefore, oncologists or geneticist doctors should use an acute computer program like the one we offer to foresee the results once they sequence the patients TP53 gene. On the other hand, our model is not only authentic for p53 protein. There are hundreds of other proteins that cause genetic diseases, and a similar model can be manufactured for each one. Furthermore, medicine designers can produce medicine that can be bound to the active part of a protein of their interest. Our artificial intelligence approach determines which part of a protein is active by categorizing mutations, thus offering a solution on this matter. When all of these aspects are considered, we can see the potential benefits and the use of our model in fields such as medicine and pharmacology.

In future, since our study can be applied to other genetic diseases based on mutation, we think that it will be recommendable to create different databases for future studies as IARC has done for p53.

## REFERENCES

[1] R. Karchin, M. Diekhans, L. Kelly, D. J. Thomas, U. Pieper, N. Eswar, D. Haussler, and A. Sali, "Ls-snp: large-scale annotation of coding non- synonymous snps based on multiple information sources," Bioinformat- ics, vol. 21, no. 12, pp. 2814–2820, 2005.

[2] D. L. Rubin, F. Shafa, D. E. Oliver, M. Hewett, and R. B. Altman, "Rep- resenting genetic sequence data for pharmacogenomics: an evolutionary approach using ontological and relational models," Bioinformatics, vol. 18, no. suppl 1, pp. S207–S215, 2002.

[3] R. H. Lathrop and M. J. Pazzani, "Combinatorial optimization in rapidly mutating drug-resistant viruses," Journal of Combinatorial Optimiza- tion, vol. 3, no. 2-3, pp. 301–320, 1999.

[4] Y. Xu, "Regulation of p53 responses by post-translational modifications," Cell death and differentiation, vol. 10, no. 4, p. 400, 2003.

[5] E. Appella and C. W. Anderson, "Post-translational modifications and activation of p53 by genotoxic stresses," European Journal of Biochemistry, vol. 268, no. 10, pp. 2764–2772, 2001.

[6] H. Qian, T. Wang, L. Naumovski, C. D. Lopez, and R. K. Brachmann, "Groups of p53 target genes involved in specific p53 downstream effects cluster into different classes of dna binding sites," Oncogene, vol. 21, no. 51, p. 7901, 2002.

[7] C. Caelles, A. Helmberg, and M. Karin, "p53-dependent apoptosis in the absence of transcriptional activation of p53-target genes." Nature, vol. 370, no. 6486, p. 220, 1994.

[8] M. Olivier, R. Eeles, M. Hollstein, M. A. Khan, C. C. Harris, and P. Hainaut, "The iarc tp53 database: new online mutation analysis and recommendations to users," Human mutation, vol. 19, no. 6, pp. 607–614, 2002.

[9] M. Mihara, S. Erster, A. Zaika, O. Petrenko, T. Chittenden, P. Pancoska, and U. M. Moll, "p53 has a direct apoptogenic role at the mitochondria," Molecular cell, vol. 11, no. 3, pp. 577–590, 2003.

[10] S. A. Danziger, S. J. Swamidass, J. Zeng, L. R. Dearth, Q. Lu, J. H. Chen, J. Cheng, V. P. Hoang, H. Saigo, R. Luo et al., "Functional census of mutation sequence spaces: the example of p53 cancer res- cue mutants," IEEE/ACM Transactions on Computational Biology and Bioinformatics, vol. 3, no. 2, pp. 114–125, 2006.

[11] A. C. Martin, A. M. Facchiano, A. L. Cuff, T. Hernandez-Boussard, M. Olivier, P. Hainaut, and J. M. Thornton, "Integrating mutation data and structural analysis of the tp53 tumor-suppressor protein," Human mutation, vol. 19, no. 2, pp. 149–164, 2002.

[12] Y. LeCun, Y. Bengio, and G. Hinton, "Deep learning," Nature, vol. 521, no. 7553, pp. 436–444, 2015.

[13] N. Rusk, "Deep learning," Nature Methods, vol. 13, no. 1, p. 35, 2016.

[14] L. Rampasek and A. Goldenberg, "Tensorflow: Biologys gateway to deep learning?" Cell systems, vol. 2, no. 1, pp. 12–14, 2016.

[15] T. Zeng, R. Li, R. Mukkamala, J. Ye, and S. Ji, "Deep convolutional neural networks for annotating gene expression patterns in the mouse brain," BMC bioinformatics, vol. 16, no. 1, p. 147, 2015.

[16] D. R. Kelley, J. Snoek, and J. L. Rinn, "Basset: learning the regu- latory code of the accessible genome with deep convolutional neural networks," Genome research, vol. 26, no. 7, pp. 990–999, 2016.

[17] H. Y. Xiong, B. Alipanahi, L. J. Lee, H. Bretschneider, D. Merico, R. K. Yuen, Y. Hua, S. Gueroussov, H. S. Najafabadi, T. R. Hughes et al., "The human splicing code reveals new insights into the genetic determinants of disease," Science, vol. 347, no. 6218, p. 1254806, 2015.

[18] B. Alipanahi, A. Delong, M. T. Weirauch, and B. J. Frey, "Predicting the sequence specificities of dna-and rna-binding proteins by deep learning," Nature biotechnology, vol. 33, no. 8, pp. 831–838, 2015.

[19] M. Lichman, "UCI machine learning repository," 2013. [Online]. Available: http://archive.ics.uci.edu/ml

[20] M.-L. Lu, F. Wikman, T. F. Orntoft, E. Charytonowicz, F. Rabbani, Z. Zhang, G. Dalbagni, K. S. Pohar, G. Yu, and C. Cordon-Cardo, "Impact of alterations affecting the p53 pathway in bladder cancer on clinical outcome, assessed by conventional and array-based methods," Clinical Cancer Research, vol. 8, no. 1, pp. 171–179, 2002.

[21] M. Abadi, A. Agarwal, P. Barham, E. Brevdo, Z. Chen, C. Citro, G. S. Corrado, A. Davis, J. Dean, M. Devin et al., "Tensorflow: Large-scale machine learning on heterogeneous distributed systems," arXiv preprint arXiv:1603.04467, 2016.

[22] F. Pedregosa, G. Varoquaux, A. Gramfort, V. Michel, B. Thirion, O. Grisel, M. Blondel, P. Prettenhofer, R. Weiss, V. Dubourg et al., "Scikit-learn: Machine learning in python," Journal of Machine Learn- ing Research, vol. 12, no. Oct, pp. 2825–2830, 2011.

[23] S. A. Danziger, R. Baronio, L. Ho, L. Hall, K. Salmon, G. W. Hatfield, P. Kaiser, and R. H. Lathrop, "Predicting positive p53 cancer rescue regions using most informative positive (mip) active learning," PLoS Comput Biol, vol. 5, no. 9, p. e1000498, 2009.

[24] L. v. d. Maaten and G. Hinton, "Visualizing data using t-sne," Journal of Machine Learning Research, vol. 9, no. Nov, pp. 2579–2605, 2008.